\documentclass[twocolumn]{svjour3MOD}
\usepackage[utf8]{inputenc}
\usepackage{amsmath,bm}
\usepackage{graphicx}
\usepackage{color}

\date{ }
\title{Simulating a burnt-bridges DNA motor with a coarse-grained DNA model}
\titlerunning{Simulating a burnt-bridges DNA motor with a coarse-grained DNA model}  

\author{Petr \v{S}ulc \and Thomas E.~Ouldridge \and Flavio Romano  \and Jonathan~P.~K. Doye \and Ard~A. Louis}
\institute{P. \v{S}ulc \and T. E.~Ouldridge  \and A. A. Louis \at
          Rudolf Peierls Centre for Theoretical Physics\\ 
          University of Oxford \\ 1 Keble Road, Oxford \\ OX1 3NP \\ United Kingdom \\
          \email{p.sulc1@physics.ox.ac.uk}\\
          \email{t.ouldridge1@physics.ox.ac.uk}\\
          \email{ard.louis@physics.ox.ac.uk } \\
          \and F. Romano  \and J.P.K. Doye \at
           Physical and Theoretical Chemistry Laboratory \\
 	   Department of Chemistry \\ University of Oxford \\ South Parks Road \\
 	   Oxford\\ OX1 3QZ \\ United Kingdom \\
 	   \email{flavio.romano@chem.ox.ac.uk} \\
	   \email{jonathan.doye@chem.ox.ac.uk}
           }

\begin{document}

\maketitle

\begin{abstract}
We apply a recently-developed coarse-grained model of DNA, designed to capture the basic physics of nanotechnological DNA systems, to the study of a `burnt-bridges' DNA motor consisting of a single-stranded cargo that steps processively along a track of single-stranded stators. We demonstrate that the model is able to simulate such a system, and investigate the sensitivity of the stepping process to the spatial separation of stators, finding that an increased distance can suppress successful steps due to the build up of unfavourable tension. The mechanism of suppression  suggests that varying the distance between stators could be used as a method for improving signal-to-noise ratios for motors that are required to make a decision at a junction of stators.

\end{abstract}

\section{Introduction}
DNA has proved itself to be a versatile material for engineering on the nanoscale. A single DNA strand consists of nucleotides covalently connected by a sugar-phosphate backbone. Nucleotides also possess one of four types of base group, adenine (A), guanine (G), cytosine (C) or thymine (T). The planar bases interact with other bases by forming parallel face-to-face stacks and hydrogen bonds along their edges, resulting in the iconic double-stranded DNA helix (dsDNA) \cite{Watson1953}. Due to their chemical structure, hydrogen-bonded C-G and A-T base pairs are the most favourable, and helices made from complementary sequences of bases form the most stable duplexes. 

The selectivity of DNA base-pairing can be used to engineer artificial
structures and devices. As was originally proposed by Seeman, sequences of
a set of single strands can be designed so that a certain configuration is
the global free-energy minimum of the system \cite{Seeman1982}. An
enormous range of nanostructures have been realised simply by cooling
solutions of single-stranded DNA (ssDNA) based on this principle.
Finite-size structures have been designed that assemble from a small number
of short oligonucleotides \cite{Goodman2005}, a single long strand and many
short staple strands \cite{Rothemund06,Douglas09} (a technique known as DNA
`origami') and a large number of short stands \cite{Wei2012,Ke2012}.
Additionally, large one-dimensional ribbons \cite{Yan2003}, 2-dimensional
arrays \cite{Winfree98,Malo2005} and 3-dimensional crystals
\cite{Zheng2009} have been realised.

DNA nanotechnology is not confined to static structures -- dynamic systems have also been realised. Duplex formation and toehold-mediated strand displacement \cite{Zhang2007} (a process in which a strand is removed from a duplex by a competing strand that can form more base pairs with the complement) allow a DNA system to respond to its environment. In particular, these processes can couple chemical change to mechanical operations and have the potential to process signals. DNA nanotweezers \cite{Yurke2000}, a switch that can be cycled through its closed and open states by the sequential addition of two types of strand, demonstrated the principle. ``Clocked'' addition of strands~\cite{Goodman2008,Andersen09,Lo2010,Han2010,Sherman2004,Shin2004} or permutation of external conditions \cite{Liedl2005,Cheng2012} have since been used in the design of a number of active systems, including some in which the mechanical change has been harnessed to induce unidirectional motion along a track \cite{Sherman2004,Shin2004,Cheng2012}. Recently, autonomous devices and walkers that function without external forcing (by catalysing the equilibration of an out-of-equilibrium system) have also been created \cite{Yin2004,Chen2004,Bath05,Tian2005,Venkataraman2007,Green2008,Bath2009,Omabegho2009,Wickham11,Muscat2011}.

The high degree of parallelisation and the ability to  interface directly with biological and molecular systems make DNA-based computation promising. In 1994, Adleman showed that DNA strands could be used to encode a Hamiltonian path problem, which was then solved upon mixing of the strands \cite{Adleman1994}. Since then, much work has  gone into developing DNA-based logic circuits \cite{Seelig2006}, with a DNA neural network that can recognise simple patterns having recently been developed \cite{Qian2011}. DNA logic has also been combined with walking devices to produce systems that can select from distinct pathways at a junction depending on solution conditions, or properties of the walker itself \cite{Muscat2011,Wickham12}.

Although DNA nanotechnology shows great promise as a field, it is far from being fully mature. In particular, optimisation of nanostructure assembly and nanodevice operation will be vital if they are to prove generally useful. DNA nanostructures and nanodevices are typically designed using well-established thermodynamic models of DNA duplex stability, such as the unified nearest-neighbour model of SantaLucia~\cite{SantaLucia2004}.
However, nanodevices and nanostructures can involve non-trivial multi-stranded complexes with pseudoknots~\cite{Dirks2007} or complex internal loops whose stabilities have not yet been incorporated into thermodynamic models. Moreover, non-equilibrium processes can be important in these systems. Furthermore, the three-dimensional structure of a DNA complex may result in tension or compression forces~\cite{Liedl2010} that cannot be described without an explicit three-dimensional representation of the system.

Computer simulations provide controllable access to time and spatial resolutions that are not accessible in experiments. Simulations of DNA nanotechnological systems therefore have the potential to offer insight into aspects of their operation and design, provided the computational model accurately describes the relevant physics.  At the greatest level of detail, DNA molecules can be studied using quantum-mechanical calculations.  However, typical of what is achievable with the most advanced methods is for example the calculation of interactions for two base pairs in vacuum~\cite{Svozil10}. Classical molecular dynamics simulation packages such as AMBER~\cite{cornell95} retain an all-atom representation of the DNA molecule and use empirical force fields between atoms. However, they too are computationally very demanding, and simulating rare events, such as breaking of base pairs in a duplex, remains at the limits of what is currently possible~\cite{Perez2012}.

Processes relevant to DNA nanotechnology typically involve reactions between multiple strands whose lengths vary from about ten to several thousands of bases. In order to address sizes and time scales at which the reactions relevant to DNA nanotechnology happen, one needs to turn to coarse-grained models of DNA \cite{Drukker2001,Sales-Pardo2005,Kenward2009,Ouldridge2009,Knotts2007,Sambriski2009,Linak11,Araque11,Florescu2011,Morriss-Andrews2010,Savin2011,Dans2010,Savelyev2009,Becker2007,Lankasbook,Svaneborg12}. Such models replace multiple atoms by a reduced set of degrees of freedom and parametrised effective interactions. Each model necessarily makes a compromise between the level of detail of its representation, physical accuracy and its efficiency, which sets its specific domain of applicability. 

In this paper we use the coarse-grained DNA model of Ouldridge et al. \cite{Ouldridge2011,Ouldridge_thesis,oxDNA} to study the operation of a `burnt-bridges' DNA motor created by the Turberfield group  \cite{Bath05,Wickham11,Wickham12}. The model, which provides a physically reasonable representation of ssDNA, dsDNA and the transition between the two, allows for simulations involving reactions of DNA systems of order tens to hundreds of bases in total on a single CPU and several thousands on a GPU. Further, it has been previously successfully applied to the study of DNA nanotechnological systems including DNA tweezers \cite{Ouldridge_tweezers_2010}, kissing hairpins \cite{Romano12a} and a two-footed DNA walker~\cite{Ouldridge_thesis}. 

We first give a brief overview of  the design of the burnt-bridges motor and the DNA model. We then describe the simulation methods used and present results of simulations of one step of the DNA walker. The motor consists of a single DNA strand (cargo) which moves from one complementary strand (stator) to the next through toehold-mediated strand displacement. We study the process for different distances between stators, different strengths of the attachments of the stators to a surface and for different lengths of the toehold, particularly focusing on the consequences of inter-stator distance.

\section{Burnt-bridges DNA motor}

\begin{figure}
\includegraphics[width=0.47\textwidth]{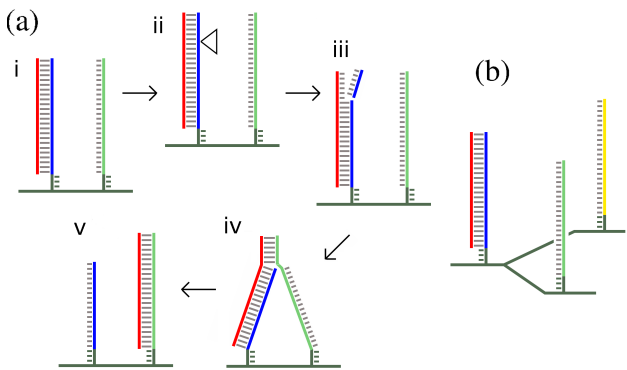}
\caption{(a) A schematic illustration of the stepping process of a burnt-bridges motor: (i) The  cargo (red) is attached to the first stator (blue), which is a complementary DNA strand. (ii) The nicking enzyme binds to its recognition sequence, and catalyses the hydrolysis of the backbone.
(iii) The nicking enzyme and stator fragment have dissociated. (iv) The exposed toehold of the cargo binds to the second stator (green). (v) The second stator fully displaces the first, and the motor completes a step. 
(b) Decision making at a junction. The cargo could step to either the yellow or green stator. By altering or blocking the toeholds, one direction or the other can be made preferable.}
\label{fig_walker}
\end{figure}

The burnt-bridges DNA motor studied in this work is a system that produces autonomous,  unidirectional motion of a single-stranded cargo along a track of single-stranded stators. The motor was experimentally realised  in Refs. \cite{Bath05,Wickham11,Wickham12}, and the stepping process of the motor is schematically illustrated in Figure~\ref{fig_walker}. The stators can be attached to a DNA duplex \cite{Bath05}, or a DNA origami surface \cite{Wickham11,Wickham12}.

Initially, the cargo strand is attached to the first stator. A nicking enzyme (N.B\,bvC\,1b) is present in the solution~\cite{Heiter2005}. These enzymes can bind specifically to a certain sequence of double stranded DNA present in the stator/cargo duplex, and cut the backbone 
of the stator strand a short distance from the $5^\prime$ end of the stator (6 bases in Refs. \cite{Wickham11,Wickham12}, and 8 in Ref. \cite{Bath05}). The binding of the shorter stator fragment is unstable at experiment conditions, and it tends to detach, revealing a short toehold on the cargo. The next available stator is positioned close to the first stator (around 7 nm in Ref. \cite{Bath05} and 6 nm in Refs. \cite{Wickham11,Wickham12}) and the exposed toehold can bind to the next stator. Strand displacement can then occur, allowing the cargo to replace bonds to the first stator with bonds to the second stator.

Once the displacement process is complete, the cargo is totally detached from the first stator and fully bound to the second one. The stepping process can now be repeated, with the stator to which the cargo is attached being cut again by the enzyme and the cargo making a step to the next stator in line. The backward step is now highly improbable, as the preceding stator has been nicked and has fewer complementary bases with the cargo strand. Used stators therefore get disabled as the cargo travels along the track,  leading to the description `burnt-bridges'. Directional motion is possible because the walker's motion catalyses the hydrolysis of the stator's backbone, a free-energetically favourable process.

In the original experiment~\cite{Bath05}, three stators were attached to a double-stranded DNA track. Fluorescence was used to demonstrate that the cargo stepped along the track, visiting the stators in order. Stators were later attached to a DNA origami surface \cite{Wickham11}, and the cargo was observed to move along a 17-stator track by atomic force microscopy. Motion along an 8-site track was also observed via fluorescence, and was strongly suppressed by the removal of a stator. Recently, motors have been designed that can choose a pathway at a junction based on information either carried by the motor itself or provided externally \cite{Wickham12}.

The suggested future applications of DNA walkers such as the burnt-bridges motor include a programmable chemical synthesis and a molecular realization of a Turing machine \cite{Bath2007}.

\section{Coarse-grained DNA model}

\begin{figure}
\centering
\includegraphics[height=4cm]{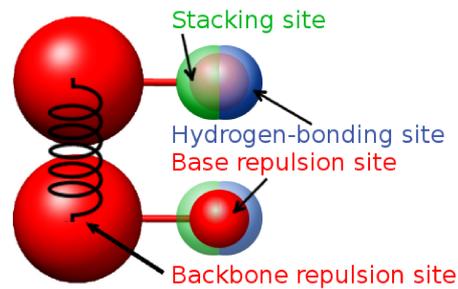}
\caption{Illustration of the structure of nucleotides in the coarse-grained model. The figure shows two nucleotides, with a spring between backbone sites to indicate the connectivity of bases within a strand (maintained by a finitely-extensible spring in the model). Each nucleotide possesses base and backbone excluded volume sites, a hydrogen bonding site and a stacking site. Red spheres indicate backbone and base excluded volume sites -- the size of the spheres shows the range of the excluded volume interaction for each site. Green and blue spheres are centred on the stacking and hydrogen-bonding sites -- the size indicates the distance at which the stacking and hydrogen-bonding interactions are maximised (when the surface of spheres from different nucleotides come into contact). Note that the interaction potentials are dependent on the relative orientation of nucleotides, as well as the location of the interaction sites.}
\label{fig_rigidbody}
\end{figure}

\begin{figure}
\centering
\includegraphics[width=0.5\textwidth]{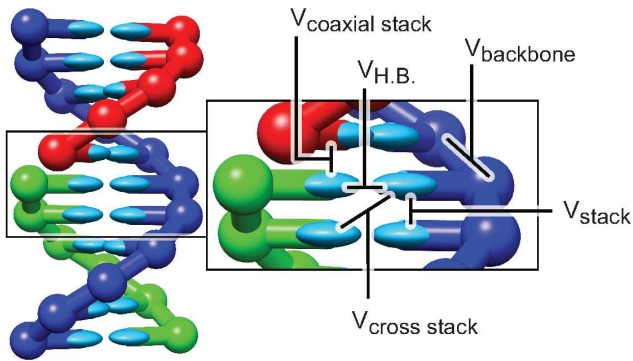}
\caption{Three strands in a double helical state with schematic illustration of the stabilising interactions. Excluded volume interactions, which are not shown, prevent the collapse of the system into a small volume, although nucleotides rarely experience it directly when in the double-helical state. In this image, and others in this article, spheres represent the backbone sites and ellipsoids the base sites. Nucleotides in the same strand are coloured identically, and the connectivity is illustrated by thin cylindrical tubes. This format of illustration is found to be clearer than the more schematic representation in Fig.~\ref{fig_rigidbody}.}
\label{fig_interactions}
\end{figure}

We use the coarse-grained DNA model introduced in Refs.~\cite{Ouldridge2011} and~\cite{Ouldridge_tweezers_2010}.
It represents a DNA strand as a string of nucleotides, each of them a rigid body with several interaction sites. The structure of the nucleotides is illustrated in Figure \ref{fig_rigidbody}.
Inter-nucleotide interactions are expressed through the system's potential
\begin{eqnarray}
 V_{\rm DNA} &= & \sum_{\left\langle ij \right\rangle} \left( V_{\rm{b.b.}} + V_{\rm{stack}} +
V^{'}_{\rm{exc}} \right) + \nonumber \\
    &+&  \sum_{i,j \notin {\left\langle ij \right\rangle}} \left( V_{\rm{HB}} +  V_{\rm{cr.st.}}  +
V_{\rm{exc}}  + V_{\rm{cx.st.}} \right) ,
 \label{eq_hamiltonian}
\end{eqnarray}
where the first sum is taken over all nucleotides that are adjacent along the backbone of a strand and the second sum is taken over all other pairs of nucleotides. The nucleotides that participate in these interactions within the duplex state are schematically shown in Figure \ref{fig_interactions}. The detailed description of the interactions along with their functional form is available in Ref.~\cite{Ouldridge_thesis}. Below we provide a basic description of each term.

The backbone potential $ V_{\rm{b.b.}}$ is an isotropic (finitely-extensible) spring potential that mimics the covalent bonds in the backbone by constraining the distance between the backbone sites of the neighbouring nucleotides within a strand. This constraint is represented through the cylindrical tubes connecting the nucleotides within a certain strand in our figures, such as Figure \ref{fig_interactions}.

Nucleotides also interact through excluded volume potentials associated with the backbone and base of each nucleotide, $V^{'}_{\rm{exc}}$ and $V_{\rm{exc}}$. The interaction radius of the excluded volume potentials is shown in Figure \ref{fig_rigidbody}. Hydrogen bonding interactions $V_{\rm{HB}}$ occur between anti-aligned complementary bases, and stacking interactions $V_{\rm{stack}}$ drive the strand to form helical stacks of bases. Together, these two interactions are the primary driving force for duplex formation. Cross stacking $V_{\rm{cr.st.}}$ is an additional secondary stabilising interaction within a duplex. Coaxial stacking $V_{\rm{cx.st.}}$ captures the stacking interaction between nucleotides that are not adjacent along the backbone of a strand. $V_{\rm{HB}}$, $V_{\rm{stack}}$, $V_{\rm{cr.st.}}$ and $V_{\rm{cx.st.}}$ are all anisotropic and depend both on the mutual orientations of nucleotides and the distances between the interaction sites.

The model was fitted at salt concentration $[\rm{Na}^{+}] = 0.5 M$, where electrostatic properties are strongly screened. Hence, for simplicity, repulsive electrostatic interactions are simply incorporated into the backbone excluded volume. Most DNA nanotechnology is indeed carried out at high salt concentration to suppress repulsion between strands. 

Although the model has sequence-specific binding (i.e., the hydrogen bonding interaction is negative only for Watson-Crick base pairs A-T and C-G), it does not otherwise distinguish between different base types in terms of interaction strengths.
The model has recently been  extended to include sequence-dependent interactions for base pairing and stacking in Ref.~\cite{oxDNA}. In this work, we use the ``average-base'' version of the model, since generic features of the system are easier to resolve without sequence-dependent complications.

The model captures the structural, mechanical and thermodynamical properties of  ssDNA and dsDNA  with average AT and GC content in the sequence \cite{Ouldridge2011}. The model was fitted to reproduce melting temperatures of short duplexes as predicted by SantaLucia's nearest-neighbour model \cite{SantaLucia1998,SantaLucia2004} for an average sequence. The single strands in the model are relatively flexible, allowing the formation of single-stranded hairpins as well as double-stranded helices.

The model's physical representation of DNA properties has allowed it  to be successfully used for the study of a wide range of DNA systems. In terms of nanotechnology, DNA nanotweezers \cite{Ouldridge_tweezers_2010}, kissing hairpins \cite{Romano12a,oxDNA} and a two-footed DNA walker \cite{Ouldridge_thesis} have been simulated. From a more biophysical perspective, the nematic ordering transition of dense solutions
of short duplexes \cite{CDMnematicDNA}, DNA overstretching \cite{Romano12b} and the extrusion of cruciform structures under torsion \cite{Matek} have also been modelled. The physically reasonable behaviour in all cases gives us confidence in applying the model to the burnt-bridges motor.

\section{Simulating the burnt-bridges motor}
\subsection{System}
\begin{figure}
\centering
\includegraphics[width=0.48\textwidth]{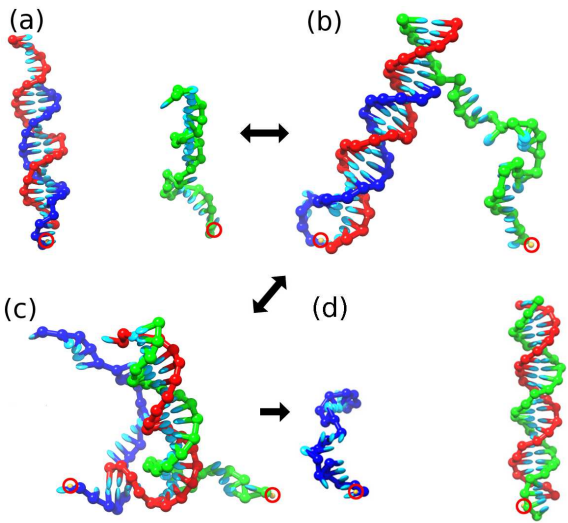}
\caption{Typical configurations that are sampled in simulations of the stepping process, with the distance between the stators $7.1\,\rm{nm}$ and attachment spring constant $131\,\rm{pN}\,\rm{nm}^{-1}$. The yellow spheres inside red circles indicate the position of the points to which the $3^\prime$ ends of the stators are attached by spring potentials. Double-headed arrows indicate transitions which are expected to be reversible under the system conditions. (a) The cargo (red) strand is attached to the first stator (blue), with a 6-base toehold exposed after the nicking of the first stator. (b) The cargo's toehold makes 6 bonds with the second stator (green). (c) Displacement is nearly complete -- the cargo has two bonds with the first stator and nineteen with the second stator. (d) Displacement is complete: the cargo is fully bound to the second stator.}
\label{fig_displacement}
\end{figure}

Our simulated system consists of three DNA strands: the first stator, the cargo strand and the second stator. The representation of these strands by the model is illustrated in Fig. \ref{fig_displacement}. We simulate the process by which the motor steps to the next stator, stages (iii)--(v) in Fig. \ref{fig_walker}. The first stator has six bases fewer than the second stator, emulating the state shown in Fig. \ref{fig_walker}\,(a)-(iii) after the small fragment of the stator has dissociated after nicking. We model the attachment of the stators to a surface via a spring potential, acting on the first $3^\prime$ base of the stators:
\begin{equation}
\label{eq_springs}
 V_{\rm{spring}}(k,\mathbf{r}_1, \mathbf{r}_2) = \frac{k}{2}\left(\mathbf{r}_1 - \mathbf{r}^0_1 \right)^2 + \frac{k}{2} \left( \mathbf{r}_2 - \mathbf{r}^0_2 \right)^2,
\end{equation}
where ${\bf r}_1$ and ${\bf r}_2$ are the positions of the centre of mass of the first $3^\prime$ nucleotide of the first and second stator respectively and ${\bf r}^0_1$ and ${\bf r}^0_2$ are the positions to which they are attached by the springs. This potential energy is included in the total potential energy of the system. We define a set of coordinates such that the stator attachment points lie in the $z=0$ plane, and are separated by a distance $d$ in the $x$ direction. We will compare three different values of $d$: $3.3$\,nm, $7.1$\,nm and $9.4$\,nm. $7.1$\,nm approximately corresponds to the distance used in Ref.~\cite{Bath05} and we chose $3.3$\,nm and $9.4$\,nm to test shorter and longer distances respectively.
 
In most simulations, we use the spring constant $k = 131\,\rm{pN}\,\rm{nm}^{-1}$, chosen so that the variance of the distance between the attached nucleotides of the first and second stators, $\left\langle |{\bf r}_1 - {\bf r}_2|^2 \right\rangle - \left\langle |{\bf r}_1 - {\bf r}_2| \right\rangle^2$, is approximately equal to the variance of the distance between two nucleotides that are 11 base pairs away on a strand in DNA duplex as simulated with our model at temperature $37\,^{\circ}{\rm C}$. Choosing the spring constant in this manner means that its magnitude is physically sensible for a DNA-based system; we will, however, consider the consequences of varying it.  

To further mimic the presence of the DNA origami substrate, we forbid the cargo and stator strands from crossing the $z=0$ plane. To achieve this we introduce an additional potential 
\begin{equation}
 V^i_{\rm{repulsion}}(\mathbf{r}_i) = 
 \begin{cases}
 \frac{k_r}{2} z_i^2, & \mbox{ if } z_i < 0 \\  0 & \mbox{  if  } z_i \geq 0  
 \end{cases}
\end{equation}
which acts on the positions of centre of mass $\mathbf{r}_i = \left(x_i, y_i, z_i \right)$ of all nucleotides $i$ in the simulation. We set $k_r$ to $1142\,\rm{pN}\,\rm{nm}^{-1}$, which is sufficiently large to effectively prevent nucleotides from crossing the $z=0$ plane. The total potential energy of the simulated system is then $V(k,d) = V_{\rm DNA} + V_{\rm{spring}}(k,d) + \sum_i^N V^i_{\rm{repulsion}} $. All our simulations are done at temperature $37\,^{\circ}{\rm C}$, the temperature used in experiment.

We use the following sequences for the first stator, second stator and cargo, respectively: 
\begin{verbatim}
5'-TCAGCCCAACTAACATTTTA-3' 
\end{verbatim}

\begin{verbatim}
5'-GGAACCTCAGCCCAACTAACATTTTA-3' 
\end{verbatim}

\begin{verbatim}
5'-CGATGTTAGTTGGGCTGAGGTTCC-3' 
\end{verbatim}
which correspond to the sequences used in Ref. \cite{Wickham11}. In the experiments, there is an additional 20-base segment at the 5$^\prime$ end of the stator that was used to bind a blocking strand that was displaced before the beginning of the stepping measurements. We do not include this segment in our simulation. The last 4 bases at the 3$^\prime$ end of the stators are not complementary to the cargo strand, acting as a flexible linker with the surface. Note that the two bases at the $5^\prime$ end of the cargo strand are not complementary to the stator strands: in the experiment~\cite{Wickham11} this dangling end was used to attach a fluorophore to the system for tracking.

\subsection{Simulation methods}

Unbiased Monte Carlo methods sample different configurations of the system with a relative probability given by the physically appropriate Boltzmann distribution ($\exp (-V/k_{\rm B}T)$, with $T$ being the temperature, $k_{\rm B}$ the Boltzmann constant and $V$ the potential energy of the configuration).
Properties of a system at a certain temperature can then be measured by averaging over the configurations obtained.
We use the Virtual Move Monte Carlo (VMMC) algorithm \cite{Whitelam2007,Whitelam2009} (specifically, the variant in the appendix of Ref. \cite{Whitelam2009}). VMMC is an approach in which new configurations are proposed by moving dynamically selected clusters of particles (nucleotides in our case), accelerating the equilibration of strongly-interacting systems relative to simple implementations in which only a single particle is moved at each step. The proposed moves used in our simulations are either rotation around a randomly chosen nucleotide's backbone site or linear translation. 

In our simulations we compute the equilibrium free energy of the system as a function of the number of base pairs between the cargo and the first stator ($b_{1c}$) and the cargo and the second stator ($b_{2c}$). We define a base pair as being formed if the hydrogen-bonding energy ($V_{\rm HB}$ in Eq.~\eqref{eq_hamiltonian}) between the two bases is more negative than $-0.97 k_{\rm B}T$, about 15\% of typical hydrogen-bonding energies in our model. The free energy $F$ of a state is related to the probability $P$ that the system is found in such a state during simulations by
\begin{equation}
F \left(b_{1c},b_{2c} \right)/ k_{\rm B} T = -\log P\left(b_{1c}, b_{2c} \right) + \log P_0\:,
\end{equation}
where $P_0$ is an arbitrary normalisation. A relatively high free energy thus corresponds to a relatively unlikely state. Free-energy landscapes do not completely determine the kinetic properties of a system, but they are indicative of how the system responds to certain changes of parameters. For example, the rate at which a process occurs is often limited by the need to pass through a high free-energy (improbable) state. Raising (or lowering) this barrier by perturbing the system generally results in an exponential decrease (or increase) in the rate of the process.

Sampling all relevant states can be difficult even with an efficient algorithm such as VMMC, as (meta)stable states are often separated by high free-energy barriers that are difficult to cross. Simulations can therefore get stuck for large portions of the simulation 
time in a local free-energy minimum. To overcome this problem, we use the Umbrella Sampling method \cite{Torrie1977}. Instead of sampling from the Boltzmann distribution, we sample from configurations using the weight $w(b_{1c},b_{2c}) \exp (-V/k_{\rm B}T)$, where $w(b_{1c},b_{2c})$ is an arbitrary biasing potential. $w(b_{1c},b_{2c})$ can be chosen to raise the probability of visiting unlikely transition intermediates, thereby accelerating equilibration. To obtain an unbiased estimate of the free energy, one must correct for the applied bias:
\begin{equation}
F \left(b_{1c},b_{2c} \right)/ k_{\rm B} T = -\log \left( \frac{P_{\rm biased} \left( b_{1c}, b_{2c} \right)}{w\left( b_{1c}, b_{2c} \right)}\right) + \log P_0,
\end{equation}
where $P_{\rm biased} \left( b_{1c}, b_{2c} \right)$ is the probability with which states appear in the biased simulation. In this work, we run several simulations to determine $w(b_{1c},b_{2c})$ by hand to enhance sampling of the relevant states. Once a satisfactory $w(b_{1c},b_{2c})$ are obtained, long simulations are run to collect the final data.

It is convenient in our case to split the Umbrella Sampling protocol into two windows. In the first window we study attachment of the cargo's toehold to the second stator, and in the second we consider the displacement process. To do this, we restrict the system to $b_{1c} \geq 14$ and $b_{2c} \leq 8$ in the first case and  $ 14 \leq b_{1c} + b_{2c} \leq 23$, $b_{1c} \geq 1$ and $b_{2c} \geq 5$ in the second. The windows are then combined using the overlap between the two with the weighted histogram analysis method \cite{Kumar1992}. 

We note that we do not sample all possible values of $b_{1c}$ and  $b_{2c}$ using these two windows. To do so would have been computationally costly without being likely to provide much insight. In particular, we do not consider the breaking of many base pairs of the first stator/cargo duplex unless the cargo is attached to the second stator. The high free-energy cost of spontaneous breaking of base pairs at this temperature make substantial melting extremely unlikely. For the same reason, we do not consider displacement intermediates with fewer than 14 base pairs involving the cargo. Finally, we do not sample the transition to states with $ b_{1c}=0$, when the cargo is detached from the first stator. To do so would be especially difficult as it would require simulating the reattachment of the cargo to the first stator in the absence of a toehold. Once detached from the first stator, however, the system is in a configuration very similar to the initial one, the only difference being the extra six base pairs available with the second stator. The free-energy change associated with adding a base pair to an isolated duplex in our model is known to be around $2.3\,k_{\rm B}T$ at $37^\circ$C (this can be seen from the slope of the free-energy profile in Fig. \ref{fig_d8}\,(b) for the initial stages of toehold binding), and so we can roughly estimate the free energies of the $b_{1c}=0$ states from those of the $b_{2c}=0$ states. Most importantly, however, only states with both $b_{1c}>0$ and $b_{2c}>0$ are expected to be sensitive to the separation of the stators, as only in these states is the cargo stretched between both.

The Umbrella Sampling simulations reported in this work involve at least $3 \times 10^{11}$ attempted VMMC moves. Half of the moves are rotations (about a random axis with an angle drawn from a normal distribution with mean zero and standard deviation 0.2 radians), and half are translations (through a random vector  with a  length drawn from a normal distribution with mean zero and standard deviation 0.85\,\AA).
 
\section{Simulation results}
We first present the free-energy landscape, and a profile along a one-dimensional pathway, for two stators $7.1\,\rm{nm}$ apart attached by springs with stiffness $k = 131\,\rm{pN/nm}$, to illustrate the basic features of the attachment of the cargo to the second stator and of the branch-migration process. 
We then compare the free-energy profiles for different attachment spring constants, different distances between the stators, and different
lengths of the toehold on the first stator.

\subsection{Stators separated by 7.1\,nm}
\begin{figure}
\centering
\includegraphics[width=0.48\textwidth]{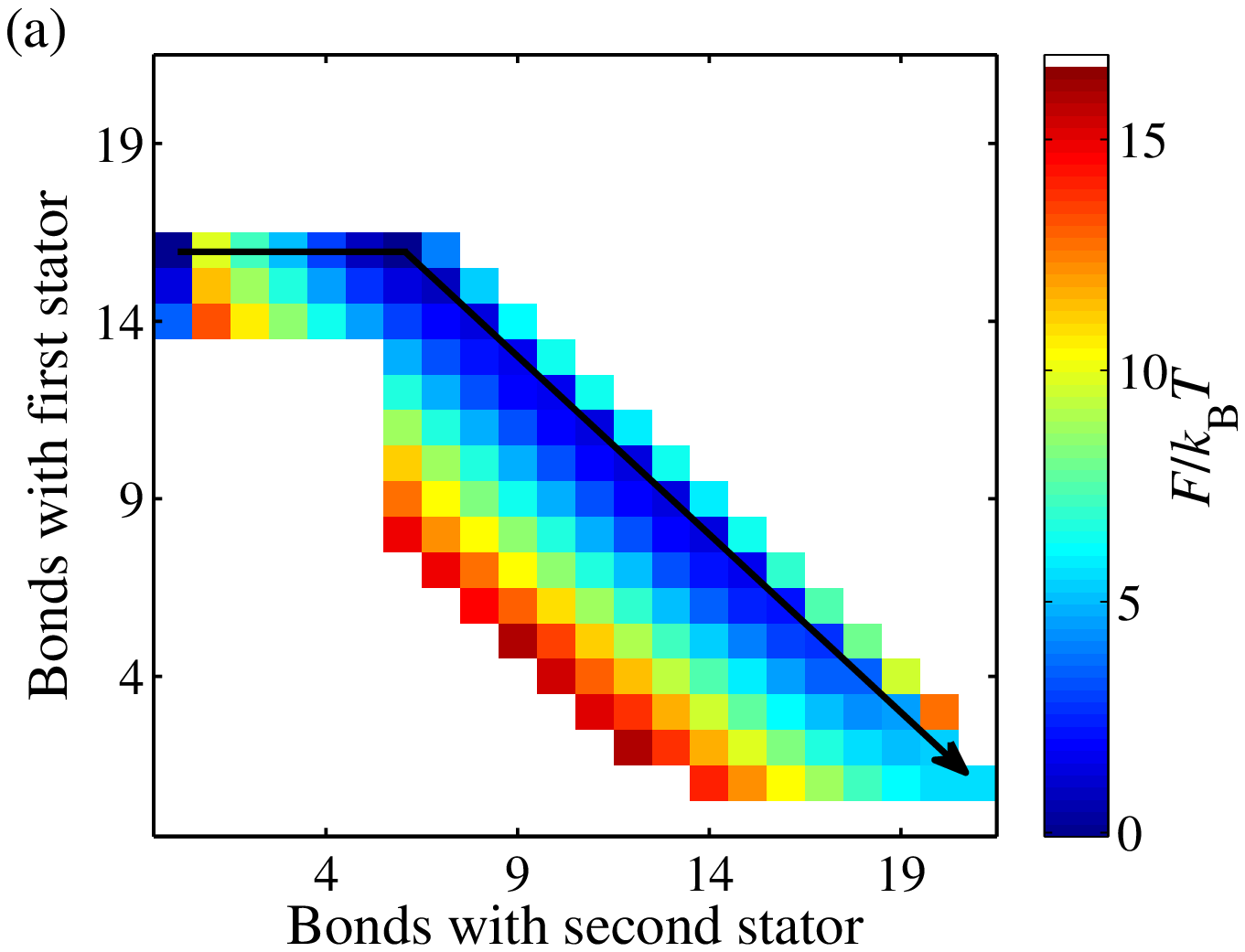}
\includegraphics[width=0.48\textwidth]{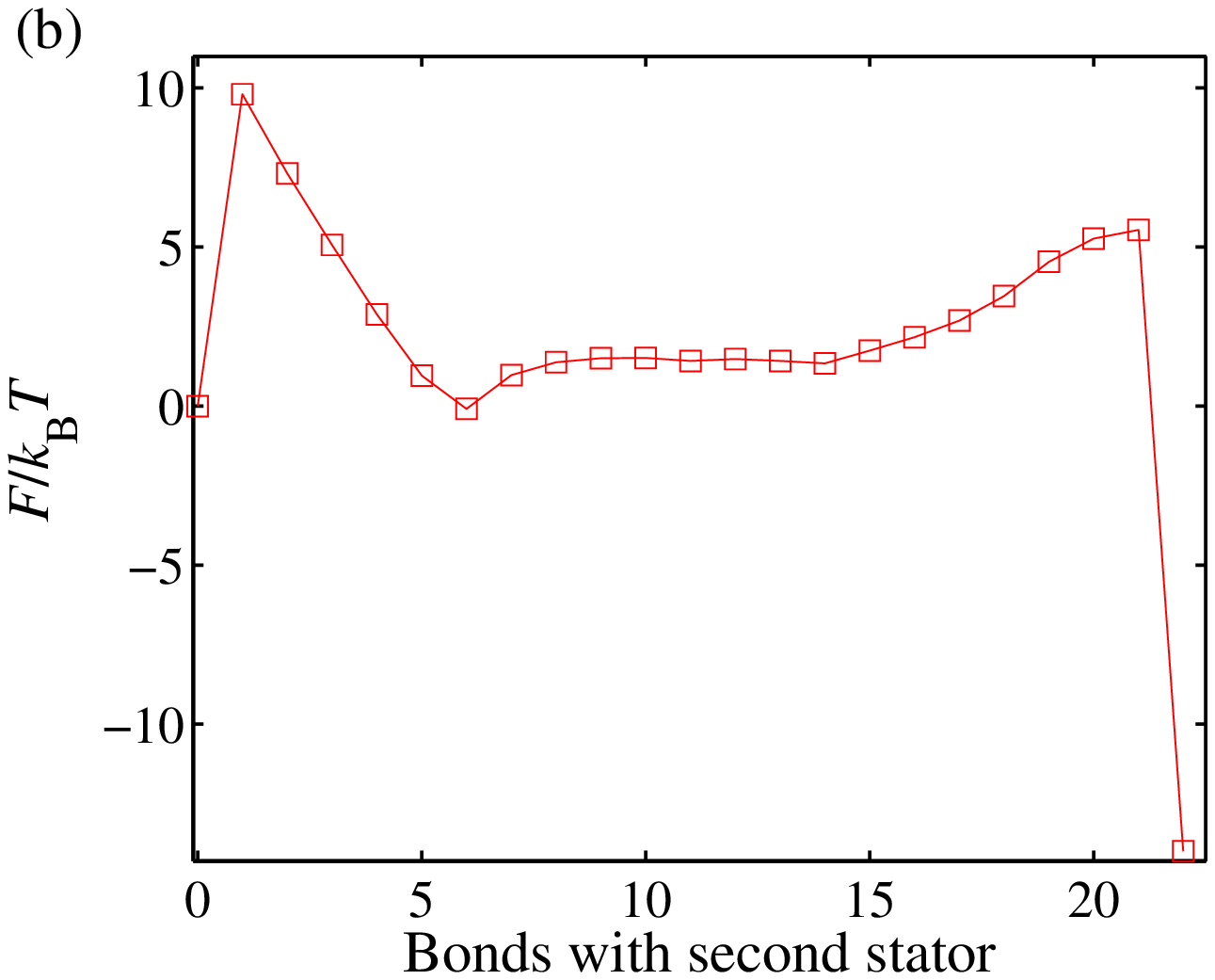}
\caption{(a) The free-energy landscape of motor stepping as a function of bonds between the cargo and the first and second stator. The arrow indicates a pathway through the landscape that is used to  plot (b). (b) The free-energy profile of displacement, plotted along the one-dimensional path shown in (a). A stage on this path is uniquely specified by the number of bonds with the second stator. The final point (23 bonds with the second stator) is estimated as discussed in the text, not measured.  }
\label{fig_d8}
\end{figure}

The free-energy landscape as a function of $b_{1c}$ and $b_{2c}$ is shown in  Fig. \ref{fig_d8}\,(a), with major features highlighted by inspecting the one-dimensional pathway shown in  Fig. \ref{fig_d8}\,(b). The free energy is normalised to be equal to $0$ for the case when the cargo has no bonds with the second stator. The basic features of the landscape are the following.
\begin{itemize}
\item There is a rise in free energy associated with the formation of the initial base pair with the second stator, due to the loss of configurational entropy when the first contact is formed.
\item The free energy decreases as successive base pairs are formed in the toehold, due to the cooperative nature of duplex formation (once the first contact is formed, successive base pairs are much more likely).
\item As displacement begins (once the seventh base pair is formed with the second stator), there is an initial rise in free energy, followed by a plateau. This initial rise  is a generic feature of displacement resulting from steric interference at the displacement interface, and is the subject of a forthcoming paper \cite{Srinivas2012}.
\item Later stages of displacement, after around 15 base pairs have formed between the cargo and the second stator, involve an increase in free energy of around $4\,k_{\rm B}T$.
\end{itemize}

An unusual feature of the free-energy profile shown in Fig. \ref{fig_d8}\,(b) is the increase in free energy towards the end of displacement. We attribute this to an increase in tension within the system. When the first bond between the cargo and the second stator is formed, the contact point is far away from the nucleotides that are attached to the surface (the $3^\prime$ end of the stators). It is therefore not difficult for the strands to reach each other at the contact point. By contrast, when more bonds have formed between the cargo and the second stator, the contact point is closer to the $3^\prime$ end of the stators. Eventually, the length of DNA between the contact point and the surface attachments gets so short that maintaing the structure causes considerable tension, which is free-energetically unfavourable and results in the observed rise in the profile.

\begin{figure}
\centering
\includegraphics[width=0.48\textwidth]{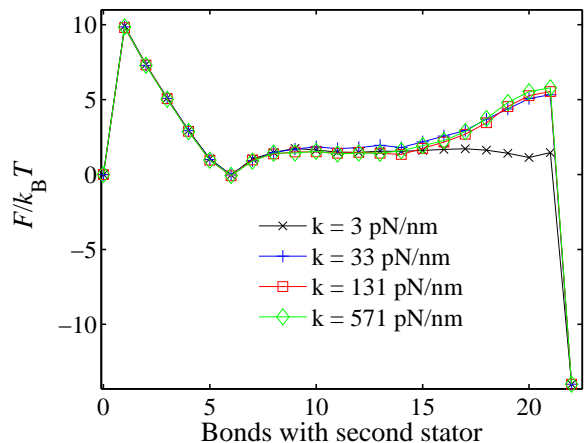}
\caption{The free-energy profile of motor stepping for various strengths of attachment to the substrate, all using a stator separation of $7.1 \rm{nm}$. The profile is taken along the path illustrated in Fig \ref{fig_d8}\,(a).}
\label{fig_multik}
\end{figure}

The role of the attachment of stators to a surface can be tested by changing the spring constant of the attachment to the surface. The results of otherwise identical simulations with different spring constants are shown in Fig. \ref{fig_multik}. Increasing or decreasing the spring constant from $k = 131\,\rm{pN/nm}$ by a factor of about 4 has a very small effect on the free-energy profile. The reason is that in this range the attachment spring is fairly stiff, and the less costly way for the system to come close together is to stretch the single-stranded sections. By decreasing the spring constant by nearly two orders of magnitude, down to $k = 3\,\rm{pN/nm}$, we are able to access a regime where the attachment springs are sufficiently weak that the strands can relax the tension effectively by moving the bases at the $3^\prime$ end of the stators closer together rather than stretching the single-stranded sections. We stress that the physically relevant regime for stators attached to a DNA duplex or to a DNA origami is the high-$k$ one.

\subsection{Varying the distance between stators}
\begin{figure}
\centering
\includegraphics[width=0.48\textwidth]{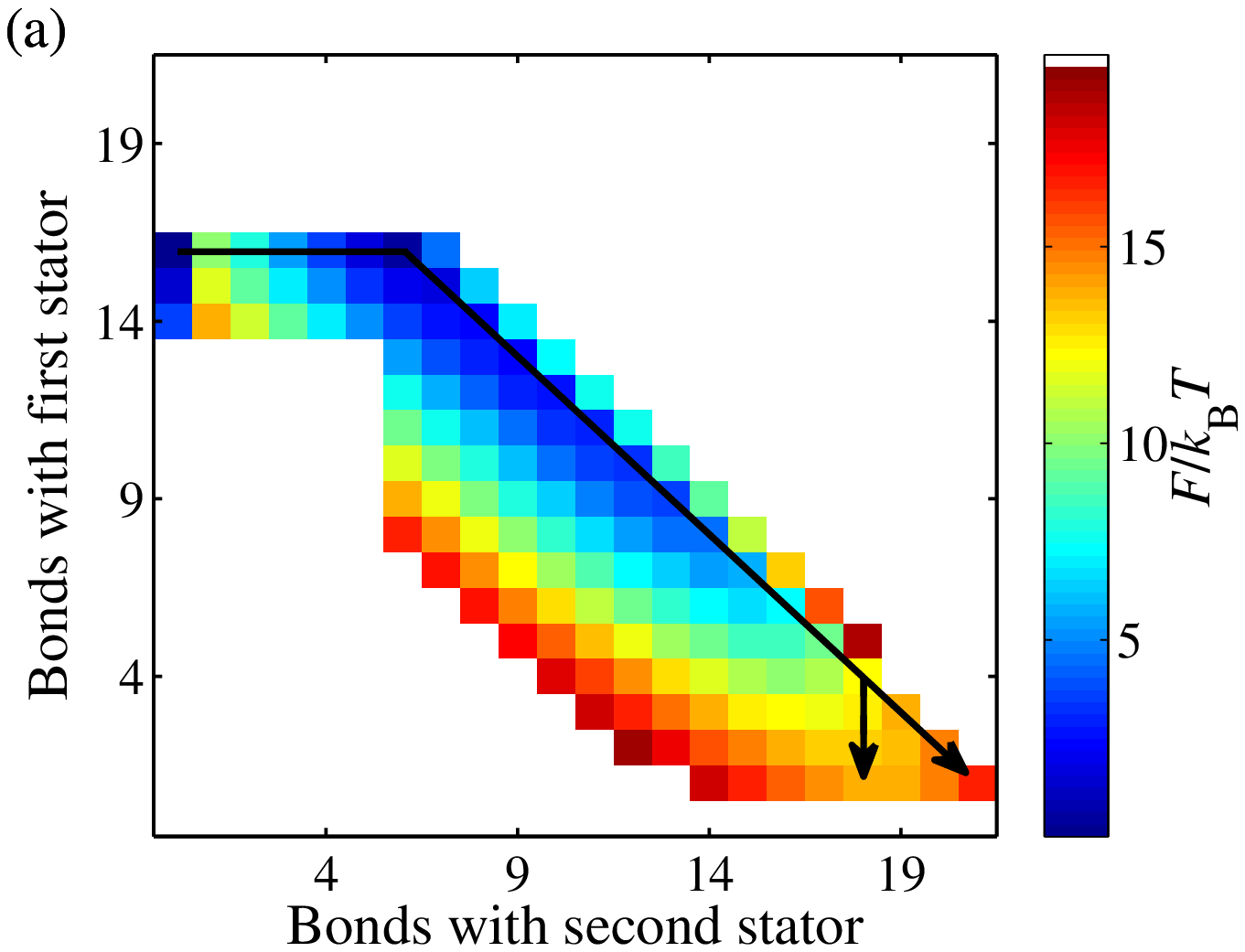}
\includegraphics[width=0.48\textwidth]{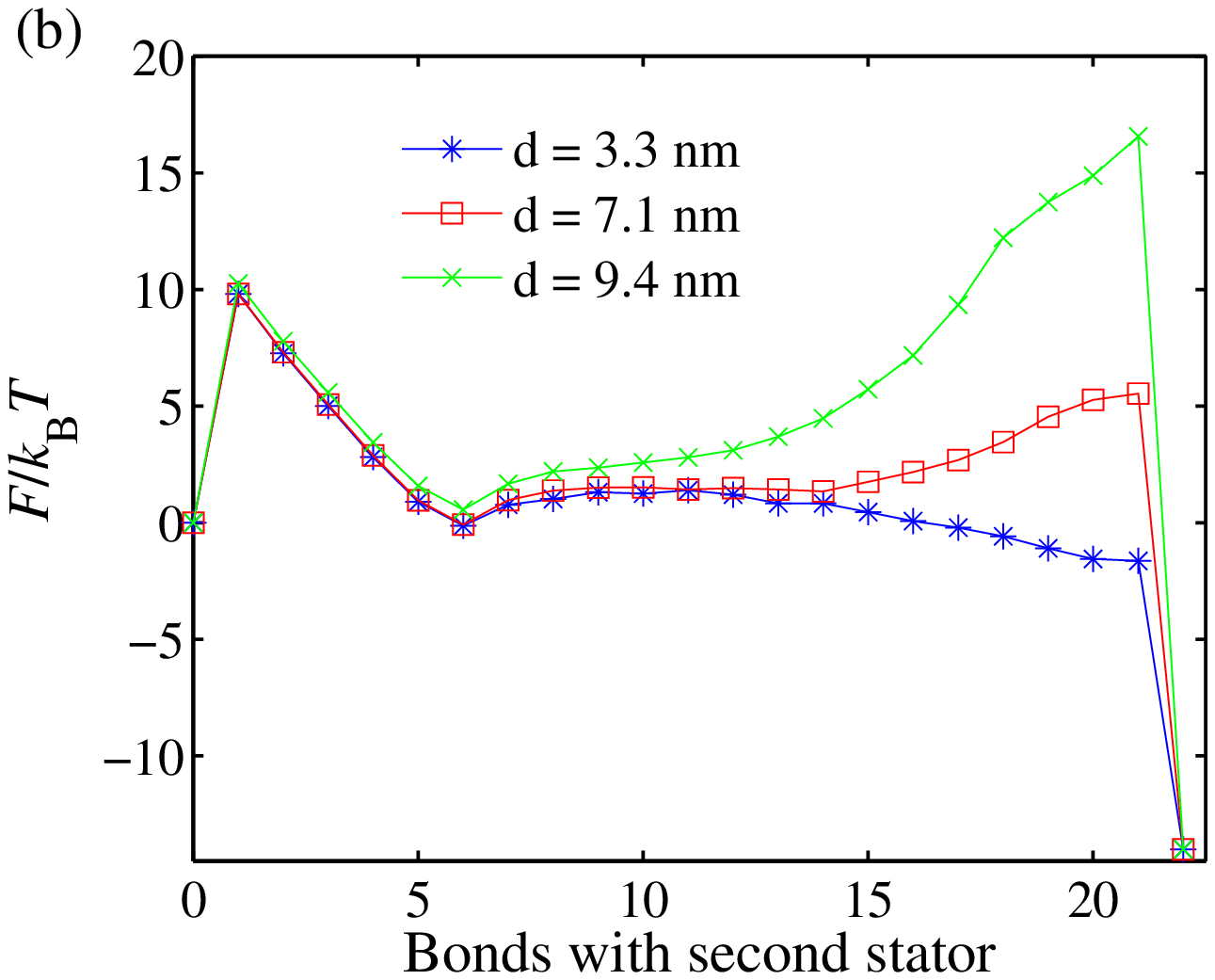}
\caption{(a) Two-dimensional free-energy landscape of stepping for $d=9.4$\,nm. (b) The free-energy profiles of motor stepping along the pathway illustrated in Fig \ref{fig_d8}\,(a) for three different distances $d$ between the attachment points of the stators.  
The illustrated alternative pathway in (a) shows that the profile in (b) for $d=9.4$\,nm probably overstates the difficulty of displacement, although it is still far more difficult than for $d=7.1$\,nm.}
\label{fig_multid}
\end{figure}

\begin{figure}
\centering
\includegraphics[width=0.48\textwidth]{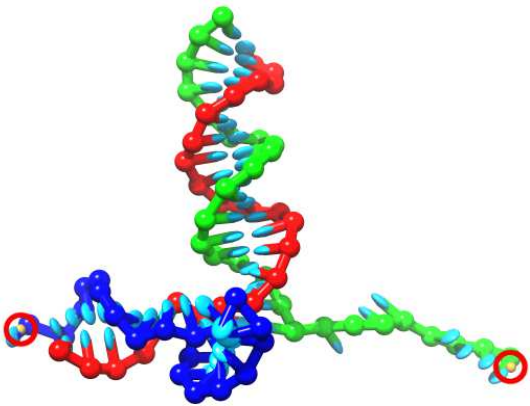}
\caption{One of the typical configurations sampled in the simulation for distance between stators $d = 9.4\,\rm{nm}$. The cargo has three bonds with the first stator and eighteen bonds with the second stator.
The tension acting on the DNA between the attachment points and the displacement interface can be clearly seen from the stretched arrangement of the second stator (green) close to its attachment point.}
\label{fig_stretch}
\end{figure}

\begin{figure}
\centering
\includegraphics[width=0.48\textwidth]{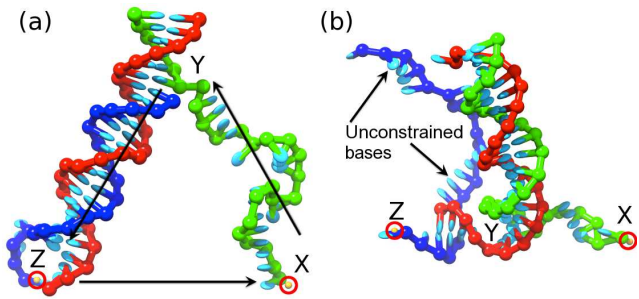}
\caption{Illustration of the reduction in constrained bases as simulation progresses. (a) Toehold-binding only: the DNA along the path $X \rightarrow Y \rightarrow Z$ is constrained by the need to for the system to from  a closed loop $X \rightarrow Y \rightarrow Z \rightarrow X$ (with the vector between attachment sites as part of the loop). (b) Later on during displacement, fewer bases are constrained within the loop. Those that remain, however, are under tension due to the need to stretch between attachment sites.}
\label{fig_loop}
\end{figure}

We now compare the stepping of the motor for three different distances between the stators: 
$d = 3.3$\,nm, $7.1$\,nm, and $9.4$\,nm. The free-energy profile along the one-dimensional pathway indicated in Fig.~\ref{fig_d8}\,(a) is shown in Fig.~\ref{fig_multid}\,(b) for all three distances. Fig. \ref{fig_multid}\,(a) shows the full two-dimensional free-energy landscape for the case of $d=9.4$\,nm. All the profiles were produced with the same spring constant, namely $k= 131\,\rm{pN}\,\rm{nm}^{-1}$. The free energies have been normalised to zero when there are no bonds between the cargo and the second stator.
One of the typical configurations sampled by our simulations for distance $9.4\,\rm{nm}$ between stators is illustrated in Fig.~\ref{fig_stretch}. 

The initial part of the free-energy profile, from 0 bonds to 6 bonds with the second stator, is nearly identical for all three distances considered. As the second stator makes more bonds with the cargo strand, we see that the increase in free energy is bigger for larger distances $d$ between the stators. This effect is consistent with our understanding that the rise in free energy is associated with increasing tension within the complex due to the need to stretch DNA between the surface attachment points and the junction. The snapshot in Fig.~\ref{fig_stretch} clearly illustrates the tension in the system at later stages of displacement for $d=9.4$\,nm.

In fact, inspection of Fig.~\ref{fig_multid}\,(a)  shows that the tension for the $9.4$\,nm case is so great that when the cargo has only one bond with the first stator, it is thermodynamically more favourable for the cargo to be bound by only 18 or 19 base pairs to the second stator, rather than by 21 base pairs as for the other values of $d$. It is therefore highly probable that a typical displacement pathway would involve the first stator detaching when the second stator has significantly fewer than 21 base pairs with the cargo, along an alternative pathway such as that shown in Fig.~\ref{fig_multid}\,(a). As such, the profile along the pathway shown in Fig.~\ref{fig_multid}\,(b) tends to overstate the difficulty in stepping between the two stators for $d=9.4$\,nm, although it is still a much more difficult process than for the shorter values of $d$.

Interestingly, for the smallest separation of $d = 3.3$\,nm there is actually a decrease in free energy  with increased binding of the cargo to the second stator.  We attribute this to the fact that the number of bases under constraint actually decreases as displacement proceeds. As can be seen from Fig. \ref{fig_loop}\,(a), when the second stator is bound only to the toehold of the cargo, most of the bases in the system are effectively held within a closed loop ($X \rightarrow Y \rightarrow Z \rightarrow X$ in the diagram -- this loop includes the vector between attachment sites) that reduces their conformational freedom. As displacement proceeds, a single-stranded section of the first stator that is not constrained by looping is generated, and the number of bases in the loop is reduced, as depicted in Fig. \ref{fig_loop}\,(b). All else being equal, transferring bases from within the constrained loop to an unconstrained section should be a favourable process because of the increase in a configurational entropy. However, the extra tension felt by the short loop overwhelms this effect for large stator separations.

\subsection{Different toehold lengths}
\begin{figure}
\centering
\includegraphics[width=0.48\textwidth]{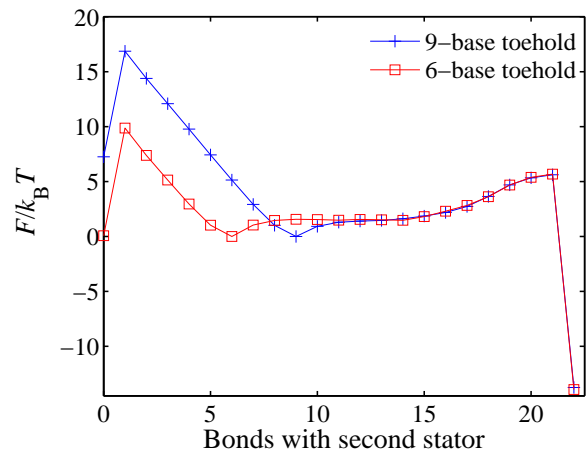}
\caption{The free-energy profile of  motor stepping for two different values of cargo toehold length. In both cases, the stators are at the same distance of $7.1$ nm. For the 6-base toehold, the profile is obtained along the path shown in Fig. \ref{fig_d8}\,(a). For the 9-base toehold, an analogous pathway in which the cargo first binds to the second stator by 9 bases and then the second stator displaces the first base-by-base is used. Once again, the final state is estimated as discussed in the text.}
\label{fig_diffnick}
\end{figure}
Finally, we compare the free-energy profile obtained at $d=7.1$ nm and $k = 131\,\rm{pN}\,\rm{nm}^{-1}$ with that for an identical system, except with a longer toehold (9 bases rather than 6) exposed by the nicking of the first stator. The free-energy profiles (taken along the same pathway as indicated in Fig. \ref{fig_d8}\,(a) for the 6-base toehold and along an analogous path for the 9-base case) are shown in Figure \ref{fig_diffnick}. These profiles have been normalised to 0 when the cargo strand is bonded fully by its toehold to the second stator, i.e. by 6 and 9 bases respectively. As expected, in the case of the 9-base toehold, we observe a larger barrier for detachment of the cargo strand from the second stator once it is bound by the toehold due to the extra base pairs. By contrast, once displacement is well underway the free-energy profiles do not differ by more 
than $0.3 k_{\rm B} T$, which is comparable with the estimated errors. Thus, once displacement has been initiated, the free-energy changes are only dependent on the distance between the strands and not the toehold length.

\subsection{Consequences of free-energy profiles for motor operation and track design}

Our data suggest that the tension generated within the motor as displacement proceeds has a potentially significant effect on motor operation. In particular, the resultant rise in free energy will suppress the speed with which the second stator can fully displace the first after binding to the toehold of the cargo. If this suppression is strong enough, the probability that the second stator detaches from the toehold rather than completing the displacement increases. The slope of the free-energy profile changes rapidly with $d$, being fairly gentle at $d=7.1$\,nm and very significant at $d=9.4$\,nm. We would therefore expect its effects on stepping success to become noticeable for $d \sim 8$ or $9$\,nm for a 6-base toehold. By contrast, we note that the initial toehold contacts have almost equal free energies for the three distances studied. So, even though the distances vary, this result suggest that initial binding rates are relatively similar in all cases. 

If stepping to the next stator after the current one has been cut is the limiting stage in motor operation, then the reduction of this rate would be manifested in the overall stepping speed of the motor. However, the binding, action and unbinding  of the nicking enzyme all contribute to the time required to take a step, so (at least at low enzyme concentrations and moderate values of $d$) this change may have a negligible effect on the overall stepping speed. At sufficiently large $d$, however, increasing $d$ further should have a noticeable effect on overall motor speed. This limit was clearly reached in Ref. \cite{Wickham11}, when removal of a single stator (estimated gap $\sim 12$\,nm) dramatically reduced the overall  rate at which the cargo reached the end of the track. Our results suggest this reduction was primarily due to the increased difficulty in completing displacement, rather than a reduced rate in making contact between the second stator and the toehold of the cargo. 

Interestingly, the potential reduction in the success probability of displacement may be advantageous at decision-making junctions such as those illustrated in Fig. \ref{fig_walker}\,(b). We might encourage the cargo to choose stator $A$ by relatively destabilising the toehold of $B$ in some manner, such as blocking it with another strand \cite{Wickham12} or by having a different sequence.  Leak currents will always exist, however, and even blunt-ended strand displacement (with no toehold) can occur \cite{Zhang2009}. 

The selection ratio of stator $A$ to stator $B$, $\phi_{A/B} $, can be modelled by assuming that the toehold of the cargo can initially bind to stator $i$ with a rate $\gamma_i$, whereupon the cargo successfully completes the step to stator $i$ with probability $f_i$ and detaches with probability $1-f_i$. In this case,
\begin{equation}
\phi_{A/B} = \frac{\gamma_A f_A}{\gamma_B f_B}.
\end{equation}
Assuming $A$ and $B$ are equidistant from the original stator, there is limited room to adjust $\gamma_A/\gamma_B$. The fact that $f_i \leq 1$  is important. It means that it will be essentially impossible to chose between two stators if both have sufficiently stable toeholds so that  $f_A$ and $f_B \sim 1$ (even if there is a large  difference in absolute stability between the two).  Furthermore, it means $\phi_{A/B}$ cannot be increased arbitrarily by  increasing the stability of toehold $A$, limiting the maximum signal-to-noise ratio.

Increasing the failure rate of displacement by adjusting $d$ (for both stators $A$ and $B$ equally) will tend to reduce $f_i$, and hence the larger toehold stability will be required to achieve $f_i \sim 1$. The junction will then be able to distinguish between toeholds that are more stable than previously, and will have a higher maximum signal-to-noise ratio. We note that if the primary consequence of increasing $d$ was to reduce the binding rate  of the toehold to the next stator (rather than the success probability once bound), it would be impossible to improve the efficiency of the junction in this way.

\section{Discussion}
We have briefly introduced a coarse-grained model of DNA, and demonstrated that
it can be used to investigate an active DNA nanotechnological device: a unidirectional molecular motor \cite{Bath05,Wickham11,Wickham12}. In particular, we have studied the physics of the stepping process from one stator to the next, once the first has been cut by a nicking enzyme, with emphasis on the dependence of this process on the separation of stators.

We observed that the free-energy profiles of initial binding of the cargo's toehold to the second stator are fairly insensitive to stator separation. However, as displacement proceeds, there is a rise in the free energy when stators are separated by larger distances, associated with the need for ever shorter sections of DNA to extend across the gap between attachment points. Such a rise will tend to reduce the speed at which the cargo completes its step to the next stator once it is bound by its toehold. For a large enough rise, there will be a reduction in the probability of successful completion of a motor's step following initial attachment. These results suggest that the experimentally observed reduction in stepping rate when the distance between stators is doubled \cite{Wickham11} is predominantly due to the increased difficulty of completing displacement, rather than a reduced probability for forming an initial contact between the next stator and the toehold of the cargo. We argue that such a reduction in successful stepping could be used to provide higher sensitivity at junctions where motors must chose between two adjacent stators. 

We find that the rigidity with which stators are held in place can be important. At low stiffnesses, stators can move towards each other and relax some of the tension generated by displacement. At high stiffnesses, the stators' movement is limited and the DNA must stretch across the full gap between attachment points instead, resulting in the aforementioned rise in free energy. These results suggest that the flexibility of the surface to which the stators are attached, and the nature of the attachment, could be significant in determining the properties of motor stepping. We estimate, however, that anchoring stators such as those studied here to a single duplex is enough to qualify as a stiff  attachment.

Our results have been obtained with a coarse-grained model, and the free-energy landscapes and profiles we have measured do not completely determine kinetics, which also depends on non-equilibrium effects. Nonetheless, the increase in tension within the motor as displacement proceeds is based on very general mechanical and structural properties of DNA that are known to be well reproduced by the model. So it is reasonable to assume that large stator separation does indeed lead to a rise in free energy with displacement progress. Furthermore, such a rise makes later  displacement intermediates harder to reach and will eventually reduce the probability of successful step completion once the second stator is bound to the cargo's toehold. For the system considered in this work, a separation of around $8$ or $9$\,nm should be sufficient to demonstrate the effects of stator separation. The most important assumption that we have made is that the small fragment of the first stator and the nicking enzyme tend to dissociate before attachment to the next stator can occur. Even if this is not the case, however, their influence will be strongest during toehold attachment and displacement initiation, and should not prevent the rise in free energy associated with increased tension at the later stages of displacement. 

Future work could include explicit simulations of the dynamics of motor stepping, although such simulations will probably be computationally demanding even for a model as simple as this one. It would also be worthwhile to  study the consequences of different sequences using the sequence-dependent model, and motifs such as internal mismatches, on the operation of the motor.

The software tools implementing our coarse-grained DNA model that were used for producing data for this work are freely available at http://dna.physics.ox.ac.uk. 


\section{Acknowledgments}
We would like to thank Shelley Wickham, Jonathan Bath, Alex Lucas and Andrew Turberfield for helpful discussions. 
The authors also acknowledge financial support from the Engineering and Physical
Sciences Research Council, University College (Oxford), and from the Oxford Supercomputing Centre for
computer time. P.~\v{S}. is grateful for the award of a Scatcherd 
European Scholarship.

\bibliographystyle{spphys}      
\bibliography{dna_biblio}

\end{document}